\documentclass{PoS}

\title{Off-shell Fragmentation}

\ShortTitle{Off-shell Fragmentation}

\author{\speaker{K. Urmossy}\\
$^{1}$Institute of physics, Jan Kochanowski University, 15 \'{S}wi\k{e}tokrzyska Street,
PL-25406 Kielce, Poland\\
E-mail: \email{karoly.uermoessy@cern.ch}}

\abstract{
A new framework is sketched for the treatment of the hadronisation of a highly-virtual quark and anti-quark jet pair created in electron-positron annihilations. As in such a case, factorization theorem does not work, a new scale-evolution equation is proposed for the fragmentation functions. In this approach, the virtuality of the leading parton (taken to be equal to the mass of the jet it induces) is used as fragmentation scale.
}

\FullConference{XXV International Workshop on Deep-Inelastic Scattering and Related Subjects\\
		3-7 April 2017\\
		University of Birmingham, UK}

\usepackage{epsfig}

\usepackage{amssymb}
\usepackage{amsmath}
\usepackage{amsfonts}
\usepackage{amsbsy}
\usepackage{amscd}
\usepackage{amsopn}
\usepackage{amstext}
\usepackage{amsxtra}
\usepackage{graphicx}
\usepackage{color}
\usepackage{slashed}
		
\newcommand{\be}{\begin{equation}}
\newcommand{\ee}[1]{\label{#1} \end{equation}}
\newcommand{\ba}{\begin{eqnarray}}
\newcommand{\ea}[1]{\label{#1} \end{eqnarray}}

\begin{document}

\section{Introduction}
In the standard approach, spectra of hadrons stemming from high-energy collision of initial objects $A$ and $B$ (which may be $e^{\pm},p,\bar p, d$ and various nuclei) are calculated as a convolution of a ``hard'' cross-section, with fragmentation (FF) and parton distribution (PDF) functions \cite{bib:Resum}. The hard cross-section is the creation of on-shell partons from the hard reaction of the initial states, while the FFs describe the hadronisation of these created on-shell partons. This is an approximation, as the partons entering and exiting the hard cross sections are intermediate particles, thus, they are off-shell. In this paper, we discuss problems that arise in a framework, in which, these partons are \textit{not} taken as on-shell. 
\begin{figure}[!h]
\begin{center}
\includegraphics[width=0.8\textwidth]{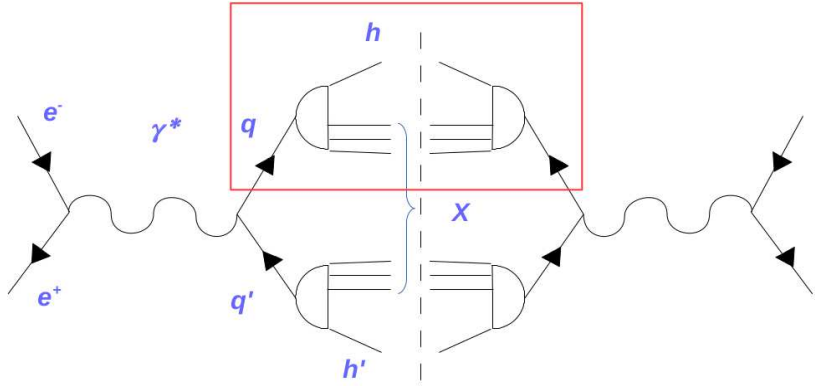} 
\end{center}
\caption{Graph of electron-positron annihilation into two jets. The subprocess of the hadronisation of a quark, described by $\mathcal{D}^h_q(p,P)$ is boxed.
\label{fig:eeqqhh2}}
\end{figure}

Let us, for example, examine the single inclusive hadron distribution in electron-positron ($e^+e^-$) annihilations with two jets in the final state (Fig.~\ref{fig:eeqqhh2}): 
\be
\frac{1}{\sigma_0} p^0\frac{d\sigma}{d^3\mathbf{p}}^{ee\rightarrow hX} \;\sim\; \frac{L_{\mu\nu} H^{\mu\nu}}{s^2} \;.
\ee{intro1}
As the leptons ($e(p_1)$ and $\bar{e}(p_2)$) are real, on-shell initial particles, the leptonic part $L_{\mu\nu} = Tr\lbrace \gamma_\mu \slashed{p}_1\gamma_\nu \slashed{p}_2\rbrace$ is the usual. However, as the quark ($q^\ast(P)$) and anti-quark ($\bar{q}^\ast(P')$) are off-shell, their propagators appear in the hadronic part 
\be
H^{\mu\nu} \;\sim\; \int d^4 P \, \mathbf{Tr}\left\lbrace \frac{1}{\slashed{P'}} \gamma^\nu \frac{1}{\slashed{P}} \mathcal{D}^h_q(p,P) \frac{1}{\slashed{P}} \gamma^\mu \frac{1}{\slashed P'} \int \frac{d^3\mathbf{p'}}{p'^0}\mathcal{D}^{h'}_{\bar q}(p',P') \right\rbrace \;,
\ee{intro2}
along with $\mathcal{D}^h_q(p,P)$, which describes the creation of hadron $h(p)$ from the virtual quark $q(P)$ (subprocess in the box in Fig.~\ref{fig:eeqqhh2}). $\mathcal{D}^h_q(p,P)$ is a $4\times4$ matrix, which contains information also on the correlations between the spin of the leading quark and the final state hadron. Averaged over the spin of the hadron, $\langle \mathcal D_q^h \rangle_{spin} \sim \mathbb{I}_4 D_q^h$, with $D_q^h$ being a function. This way, in the spin-averaged case, $H$ simplifies to
\be
H^{\mu\nu} \;\sim\;  g^{\mu\nu} \int d^4 P \, \frac{D^h_q(p,P)}{P'^2 P^2}    \int \frac{d^3\mathbf{p'}}{p'^0} D^h_q(p',P') \;. 
\ee{intro3}
As the quarks are off-shell, their momenta 
\be
P  = \left(P^0, \mathbf P\right) \;,\qquad P' = \left(\sqrt s/2 - P^0,- \mathbf P\right)\;
\ee{intro4}
are not fully determined by energy-momentum conservation (as would be in the factorised case, in which, $P,P' = \left(\sqrt s/2,\pm \sqrt s/2,\mathbf 0\right)$). As a consequence, the masses of the jets are not equal to each other, which is in accordance with experiments \cite{bib:jetmass}. As at least one pion has to be created in each jets, it is reasonable to restrict the domain of integration in Eq.~(\ref{intro3}) with respect to $P$ to the region where $P'^2, P^2 \geq m_\pi^2$.



It is important to note that it is not straightforward to give an operator definition of $\mathcal D$ using Wilson lines \cite{bib:fact}, as the total virtuality $P^2$ of the partons radiated during the branching process is \textit{not} necessarily small compared to their total energy $P^0$ or threemomentum $|\mathbf P|$: $P^2=M^2\not\ll P^0, |\mathbf P|$.

\section{Off-shell scale evolution}
As a first step, we work in the $\phi^3$ theory in 6 dimensions (which is the simplest asymptotically free field theory), and model the branching by the self-similar process depicted in Fig.~\ref{fig:qhX}. Unlike the standard approach, here, we do not regard the radiated partons as on-shell final-state particles.  
\begin{figure}[h]
\begin{center}
\includegraphics[width=0.8\textwidth]{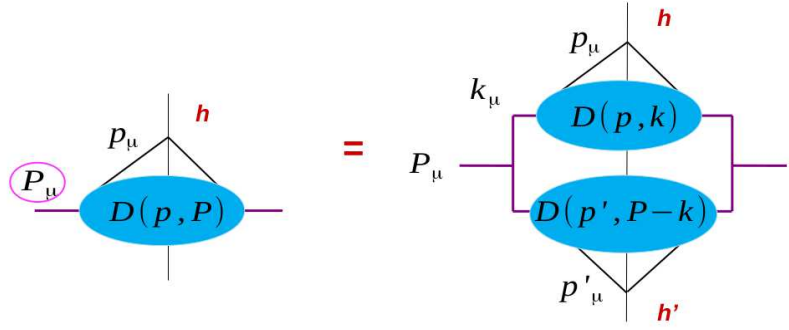} 
\end{center}
\caption{Self-similarity of the branching process inside a jet, initiated by a virtual leading parton of momentum $P$ ($P^2 = M^2 \not\approx0$) and resultng in the creation of hadrons, among which, one has momentum $p$.
\label{fig:qhXoff}}
\end{figure}
The self-similarity conjecture leads to the equation
\be
D(p,P) \;=\; \int d^6k \, \frac{D(p,k)}{ k^4 (P-k)^4}    \int \frac{d^5\mathbf{p'}}{p'^0} D(p',P-k)  
\ee{eq0}
for the hadronisation function $D(p,P)$, which we parametrise as a product
\be
D(p,P) \;=\; P^4\, \rho\left(P^2\right)\, d(p,P) 
\ee{eq01}
with a distribution $d(p,P)$  normalised as $\int (d^5p / p^0) d(p,P) = 1$ and an ``internal'' mass-distribution $\rho(M^2)$ normalised as $\int dM^2 \rho(M^2) =1$. The $P^4$ factor renders the mass dimensions of $D$. This way, Eq.~(\ref{eq0}) becomes
\be
D(p,P) \;=\; \int d^6k \, \frac{D(p,k)}{ k^4}  \rho\left[(P-k)^2\right] \;. 
\ee{eq02}
To obtain an equation for the internal mass distribution, we integrate Eq.~(\ref{eq02})  with respect to $\int (d^5p / p^0)$ and arrive at 
\be
M^4 \rho\left(M^2\right) \;=\; \int d^6k \, \rho\left(k^2\right)  \rho\left[(P-k)^2\right] \;. 
\ee{eq03}
In the present model, Eqs.~(\ref{eq02})~and~(\ref{eq03}) serve for the determination of $D(p,P)$.

In the next section, we obtain an equation similar to the Dokshitzer-Gribov-Lipatov-Alterelli-Parisi (DGLAP) \cite{bib:DGLAP} for $D(p,P)$.

\subsection{On-shell daughter partons}
\begin{figure}[h]
\begin{center}
\includegraphics[width=0.8\textwidth]{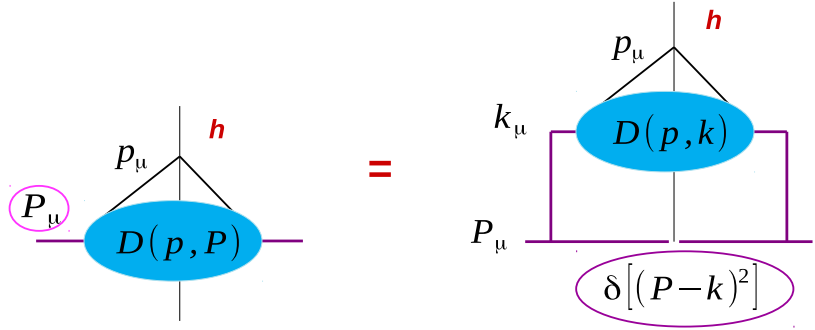} 
\end{center}
\caption{The branching process (Fig.~\ref{fig:qhXoff}) in the case, when the virtuality of the leading parton decreases step-by-step, as it radiates \textit{on-shell} daughter partons.
\label{fig:qhX}}
\end{figure}

If we push the virtuality of the daughter partons, radiated by the leading parton down to zero, we arrive at the parton-ladder shown in Fig.~\ref{fig:qhX}. This means the substitution $\rho\left[(P-k)^2\right] \rightarrow \delta\left[(P-k)^2\right]$ in Eq.~(\ref{eq02}).  
In the calculations, we use light-cone coordinates, so a general vector $k$ has the form of $k_\mu = (k_+ , k_-,\mathbf{k_T})$, with $k_\pm = (k_0\pm k_z)/\sqrt 2$. The scalar product of $a_\mu$ and $b_\mu$ is $ab = a_+b_- + a_-b_+ - \mathbf{a_T}\mathbf{b_T}$. We choose a frame comoving with the initial parton, where, its momentum is $P_\mu = (M/\sqrt 2,M/\sqrt 2,\mathbf{0} )$. The momentum of the massless final state hadron is $p_\mu = (\sqrt{2}p,0,\mathbf{0})$. Following the arguments in \cite{bib:UKFF}, we parametrize the hadronisation function as $D\left(x, Q^2 \right)$ with variables $x = 2p_\mu P^\mu / P^2$ being the energy fraction the hadron takes away from the hadron in the co-moving frame; and fragmentation scale $Q^2 = P^2$, being the initial parton's virtuality. This way, the parton ladder (Fig.~\ref{fig:qhX}) leads to the equation
\be
D\left(\frac{2p}{M}, \frac{M^2}{\Lambda^2}\right) \;=\; \kappa_4\int dk_+ \int dk_- \int d\mathbf{k_T}^2\,\mathbf{k_T}^2\, \frac{g^2(k^2 / \Lambda^2)}{k^4} \,\delta\left[(P-k)^2\right] \, D\left(\frac{2pk}{k^2}, k^2 / \Lambda^2 \right) \;.
\ee{eq5}
$\kappa_4$ is the surface of the 4-dimensional sphere, and $\Lambda$ is the scale at which the 1-loop coupling of the $\phi^3$ theory $g^2 = 1/\beta_0 \ln(Q^2/\Lambda^2)$ blows up. The $\delta$ function indicates that the emitted partons are on-shell. We imply this condition to eliminate $k_-$, and rewrite Eq.~(\ref{eq5}) with new variables $x = 2p/M$, $z = \sqrt{2}k_+/M$, $\mu = \mathbf{k_T}^2/M^2$ and $\alpha = 1- \mu/z(1-z)$. The result is
\be
D\left(x, \frac{M^2}{\Lambda^2}\right) \;=\; \frac{\kappa_4}{2} \int\limits_x^1 \frac{dz}{z} z(1-z) \int\limits_{\alpha_0}^1 d\alpha \, \frac{1-\alpha}{\alpha^2} \, g^2\left[z\frac{M^2}{\Lambda^2} \alpha \right] \; D\left[\frac{x}{z} \frac{1 - z(1-\alpha)}{\alpha}, z\frac{M^2}{\Lambda^2} \alpha \right] \;.
\ee{eq6}
The $\alpha\approx0$ region is excluded from the integral by lower cut-off $\alpha_0$, as in that region $k^2\approx0$, thus, the leading parton would become on-shell after the splitting. As it has to create at least one pion, its virtuality cannot go to zero. The $\alpha\approx1$ case is when $\mathbf{k_T}\approx0$, thus, the emission is nearly collinear.  

To obtain a DGLAP-like equation, first we set $\alpha=1$ in the first argument of $D$ under the integral. Then we change the integration variable $\alpha\rightarrow \tilde\alpha = \alpha M^2/\Lambda^2$, and differentiate with respect to $M^2/\Lambda^2$ twice (application of $\partial/\partial (M^2/\Lambda^2)$ only once would not be sufficient). As a result, we arrive at a DGLAP-like equation 
\be
\left(M^2/\Lambda^2\right)^2\left(\frac{\partial}{\partial (M^2/\Lambda^2)}\right)^2 D\left(x, \frac{M^2}{\Lambda^2}\right) \;\approx\; \frac{\kappa_4}{2} \int\limits_x^1 \frac{dz}{z} \Pi(z) \, g^2\left[z\frac{M^2}{\Lambda^2} \right] \; D\left[\frac{x}{z} , z \frac{M^2}{\Lambda^2} \right] \;
\ee{eq7}
with $\Pi(z) = z(1-z)$, which is just the leading-order (LO) splitting function in the $\phi^3$ theory, apart from the missing $\delta(1-z)$ term.

The $z$ dependence in $g^2$ and $D$ can be taken out with a simple trick used in the Modified-Leading-Log Approximation (MLLA) \cite{bib:dEnterria1}. Using $\zeta=\ln(1/z)$ and $Y=\ln (M^2/\Lambda^2)$ we can write
\ba
g^2\left[e^{-\zeta + Y} \right] \; D\left[\ast , e^{-\zeta + Y } \right]
= e^{-\zeta\, \partial/\partial Y} g^2\left[e^Y \right] \; D\left[\ast , e^Y \right] = z^{\partial/\partial Y}  g^2\left[e^Y  \right] \; D\left[\ast , e^Y \right] \;.
\ea{eq8}
This way, Eq.~(\ref{eq7}) takes the form 
\be
\partial_Y (\partial_Y -1) D\left(x, e^Y\right) \;\approx\; \frac{\kappa_4}{2} \int\limits_x^1 \frac{dz}{z} \Pi(z)  \, z^{\partial/\partial Y}  \; g^2\left[e^Y\right] \; D\left[\frac{x}{z} , e^Y\right] \;,
\ee{eq9}
which can be factorized in Mellin space:
\be
\partial_Y (\partial_Y -1) \tilde{D}\left(\omega, e^Y\right) \;\approx\; 
 \frac{\kappa_4}{2}  \; \tilde{\Pi}\left(\omega + \frac{\partial}{\partial Y} \right)  \; g^2\left[e^Y\right] \; \tilde{D}\left[\omega , e^Y \right] \;,
\ee{eq10}
with the Mellin transform of a function $f(x)$ defined as $\tilde f (\omega) = \int\limits_0^1 dx\, x^{\omega-1}f(x)$. 

\section{Remarks}
In the QCD version of the above model, substitution of the corresponding form of Eq.~(\ref{eq01}) into Eq.~(\ref{intro3}) would yield the jet mass distribution in $e^+e^-$ to two-jet events:
\be
\frac{d\sigma}{dM^2} \;\propto\; \int d^4k \, \rho\left(k^2\right)  \rho\left[(p_1+p_2-k)^2\right] \delta\left(k^2-M^2\right)\;. 
\ee{r1}

If in Eq.~(\ref{eq5}), the virtuality of the mother parton (after having emitted the daughter parton) was $k^2 = -\mathbf{k}_T^2$, we would have 
\be
\dots \int\limits^{Q^2} \frac{d\mathbf{k}^2_T \, \mathbf{k}^2_T}{k^4} \sim \int\limits^{Q^2} \frac{d\mathbf{k}^2_T \, }{\mathbf{k}^2_T }\;,
\ee{eq11}
thus, it would be enough to differentiate Eq.~(\ref{eq6}) with respect to $M^2/\Lambda^2$ only once, and we would arrive at 
\be
\partial_Y \tilde{D}\left(\omega, e^Y\right) \;\approx\;  \frac{\kappa_4}{2}  \; \tilde{\Pi}\left(\omega + \frac{\partial}{\partial Y} \right)  \; g^2\left[e^Y\right] \; \tilde{D}\left[\omega , e^Y \right] \;,
\ee{eq12}
which is the DGLAP equation in the $\phi^3$ theory (except for the lack of the $\delta(1-z)$ term in the slpitting function). The $\delta$ term in DGLAP is responsible for the proper normalisation of the probability distribution of a hadron in the jet induced by the initial parton. In the present model, $d(p,P)$ in Eq.~(\ref{eq01}) plays the role of such a probability distribution, which is normalised properly by definition.

The above model certainly lacks some sort of an operator definition for $\mathcal D$. Besides, in order to justify the self-similar approximation for the branching process (Fig.~\ref{fig:qhXoff}), it is needed to be shown that gluon exchange between the jets is suppressed by powers of a lagre scale e.g. virtuality of the leading parton.

\section*{Acknowledgement}
This work was supported by the Polish National Science Center grant 2015/19/B/ST2/00937.


\end{document}